\documentclass[useAMS,usenatbib,referee]{mn2e}

\usepackage{graphicx}
\def\dfrac#1#2{{\displaystyle#1\over\displaystyle#2}}

\title{Cosmological and kinematical criteria for the ICRF2 sources selection}
\author[Sazhin Mikhail]
{Sazhin M.V., Sementsov V.N., Zharov V.E., 
\newauthor
Kuimov K.V., Ashimbaeva N.T.,  Sazhina O.S. \\
Sternberg Astronomical Institute, Moscow State University, University pr. 13, Moscow, RUSSIA
  }
\date{Accepted ;
      Received ;
      in original form }
\pagerange{\pageref{firstpage}--\pageref{lastpage}} \pubyear{2009}

\begin{document}

\maketitle

\label{firstpage}

\begin{abstract}

The most precise realization of inertial reference frame in astronomy is the catalogue of 212 
defining extragalactic radiosources with coordinates obtained during VLBI observation runs in 1979-1995. IAU decided on the development of the second realization of the ICRF2  catalogue. The criteria of best sources selection (in terms of coordinates stability) must be defined as the first aim. The selected sources have to keep stable the coordinate axes of inertial astronomical frame.

Here we propose new criteria of source selection for the new ICRF catalogue. The first one we call as ``cosmological'' and the second one as ``kinematical''. The physical basis of these criteria is based on the assumption that apparent motion of quasars (at angular scale of the order of hundred microarcseconds) is connected with real motion inside quasars. Therefore apparent angular motion corresponds to real physical motion of a ``hot spot'' inside a radio source. It is shown that interval of redshift $0.8 \div 3.0$ is the most favorable in terms that physical shift inside such sources corresponds to minimal apparent angular shift of a ``hot spot''. Among  ``cosmologically'' selected sources we propose to select motionless sources and sources with linear motion which are predictable and stable over long time interval.

To select sources which satisfies such conditions we analyzed known redshifts of sources and time series obtained by our code and by different centers of analysis of VLBI data. As a result of these analyses  we select 137 sources as a basis for the ICRF2 catalogue.

\end{abstract}


\section{Introduction}\label{introduction}

The modern International Celestial Reference System (ICRS) is
based  on the positions of specially selected compact extragalactic radio sources (quasars, active galactic nuclei (AGN), and blazars). The system
axes are concerted with the axes of the FK5 system, and the system
origin is located in the solar system barycentre~(see \cite{zhar06}). The physical realization of the ICRS is the International Celestial Reference Frame (ICRF), which is realized
as catalogue of 608 reference extragalactic radio sources,~\cite{ma98}, most of them are quasars. The catalogue is based on the observations on the very long baseline
interferometers (VLBI), made for the period 1979--1995.

The  sources have been distinguished as: ``defining'', ``candidates'', and ``others''. 
It was supposed that ``defining'' sources to be pointlike objects, they are unresolvable
at the interferometric baselines with the length compared with the
Earth diameter. Their positions were determined with the best
current accuracy, the coordinate errors were of the order of 0.25
mas. Therefore, these sources are used to define the ICRF axes.
The coordinate stability of ``defining'' sources guaranteed the
stability of the orientation of the ICRF axes in the space. The
``candidates'' and ``others'' sources were included into
the catalogue for more dense filling of the celestial sphere and
also to link the ICRF to the optical catalogues.

The  catalogue of the radio sources,~\cite{ma98}, was published more than 10
years ago. During the past period it was carried out new
observations of the references radio sources. The duration of the
observations for several of them is about 30 years, therefore it was
appeared a proposal to revise the list of sources and also to
elaborate new criteria to select the best (the most stable) sources to construct new celestial reference frame.

Taking  in account all new ideas on the XXVI General Assembly IAU it
was made a decision on a new catalogue ICRF2. The main method to
realize this catalogue was based on the kinematical
principle,~\cite{Kov}. In other words, to built the reference
system we suppose the apparent motions and objects parallaxes 
to be neglected. But this assumption is not acceptable anymore.
According the modern observations,~\cite{mcm03}, the reference
quasars and other objects possess large angular velocities,
the apparent velocities of their motion can sometimes exceed the
speed of light.

There are  two reasons of such motion. The first one is that
there are some kind of motions inside the source. The second one
is that the light (radio waves) is refracted propagating from a
source to an observer. The refraction does not depend on the
wavelength. The achromatic refraction is the usual property of
gravitational fields, so we first briefly analyzed the hypothesis that the
apparent motions of the ICRF sources are due the effect of weak
gravitational microlensing by stars and dark objects in the Milky Way
(see, for instance~\cite{sazh, sazh98, sazh01, evan01}). But our
estimations showed that weak microlensing can not explain so numerous
apparent motions of quasars.

The more realistic  opportunity is that there are the motions
inside the sources. In this case the more appropriate model is the
Blandford-Rees model,~\cite{bla79, beg84}. The main idea is that
the quasars and AGN - objects represent the system of a massive
black hole and jets. The optical radiation is formed in black hole accretion
disk while the radio emission is formed into the jet, at
some distance from the optical source.

Initially the  Blandford-Rees model has been elaborated to explain
the observed superluminal motions in well resolved radio sources which
have not been included in the ICRF list. But it also successfully
works for radio sources from the ICRF list. The simulations done by our
group showed the good agreement between the model and
observational data by example on several sources from the ICRF
list,~\cite{zhar09}. The main idea of this model is that optical emission 
and radio emission are emitted from spatially separated regions. The indirect confirmation of this idea is measurements of optical positions of the ICRF sources and comparison of these positions with radio positions of the same sources~\cite{zach99, net02, ass03, ass07}. 
Though the authors
of cited papers explain these difference by systematic errors, it
is possible to assume the real effect which depends on the
structure of the ICRF sources. We will discuss it briefly.

The apparent  motions of the radio sources (quasars) occur at bounded angular scale of the order of distance between radio source and corresponding optical source. In linear scale it corresponds to several hundred ps or kps. The size of the radio source (the so-called ``hot spot'' in jet) is much smaller.  This spot can move with relativistic speed. It leads to high speed or even superluminal apparent motion of the radio source on the sky.

The spot motion is  interpreted as radio source motion, and exactly these motions cause the nonstationarity of the coordinate system based on extragalactic sources.

We estimated (~\cite{zhar09}) a range of the motions and predict them. The larger the range the larger the apparent angular motion of the radio source on the sky and therefore the less stable the coordinate system based on such sources.

Thereby  to improve the coordinate system stability we have to choose as remote sources as possible. It is correct in the euclidean space: the more remote a quasar, the less the angular scale of its apparent motion. In the Friedmann model
of expanding Universe it is not correct. Quasars are extragalactic
objects and have to be considered in the modern cosmological model
of expanding space-time.

In expanding  Universe angular scale of an object decreases as
this object is receding away from an observer. Then it reaches
some minimal size, and begins to increase again. The distance at which an
object has the minimal angular size in the Standard Cosmological Model
corresponds to the redshift $z = 1.63$. Therefore sources are
located nearby this distance have the minimal angular size of
their apparent motions.

The main aim of  this work is to produce a list of radio sources
which are optimal for formation of the most stable coordinate
system.

We use two criteria to choose the  appropriate sources.

The first one  we introduce as cosmological criterion. Let us
choose the group of sources located nearby the optimal value of
redshift $z_{opt} = 1.63$. We
consider the uniformly distributed group of sources having their
intrinsic motions at the scale $L$.  These motions account for
change of the brightness center position and therefore for
coordinates stability loss. Depending on the distance to the
source these changes lead to different angular motions. There is
the value of redshift $z_{opt} = 1.63$ when the angular motions are minimal. 
Therefore the first criterion to choose the
sources is to identify sources which belong the redshift
interval $z_{opt} - \Delta z_1 \le z \le z_{opt} + \Delta z_2 $. We
choose an interval $\Delta z_1 + \Delta z_2$ and compile a list of all sources
inside this interval.

The second criterion  is defined by the intrinsic motions in 
source. Let us consider the source as a black hole surrounding by
accretion disk and having two jets with opposite directions.
The black hole is precessing, therefore the jets are precessing
too. As a result of this process an observer sees the motion of a
''hot spot'' on the celestial sphere or variation of right ascension and declination . In addition, a cloud of interstellar
medium could penetrate into the jet (or jet to run accross a fixed cloud due to precession), that leads the cloud to be
accelerated and produced radio emission. The brightness
center of the system of the ``hot spot'' and accelerated cloud will
begin to shift.

The precession motion of the jets occurs with
small angular velocity according to the theory, that looks like a linear motion
of the ``hot spot'' or motion without acceleration for the period of the order 
of 30 years. At opposite, the motion of the cloud into the jet occurs with significant
acceleration. From the observational point of view the first type
of the motion is the motion with constant velocity $\dot \alpha
\neq 0, \ddot \alpha = 0; \; \dot \delta \neq 0, \ddot \delta =
0$. The second one is the accelerated motion, the acceleration is
$\dot \alpha \neq 0, \ddot \alpha \neq 0 ; \; \dot \delta \neq 0,
\ddot \delta \neq 0$. The motion due to precession is easy
predictable and all changes occur during time period of hundred
and thousand years. Penetration of a cloud into the jet is a
stochastic and almost unpredictable event. Therefore taking into
account the claim of stability of the source position and of the
coordinate system we have to choose the sources with linear motion
or stationary ones. It is recommended to exclude from the ICRF
list the sources which have accelerated motion time to time.

Therefore,  the second criterion is exclusion of sources with second derivatives 
and compilation the list of the ICRF2 with fixed sources or sources with linear motion.

We also compared our list of ``good'' candidates with
the list made by \cite{fei06}.

\section{Angular size of an extragalactic source}\label{size}

Accuracy of an extragalactic  source position determination is
defined by several characteristics of the source. One of the most
important is internal size or brightness distribution over the
source area. A median angular size of brightness distribution
determines roughly accuracy of location of this source. Therefore,
common prejustice is: the farther source the smaller its angular
size. This idea comes from Newtonian physics
which is not valid in the expanding Universe.

The angular size in the FRW  Universe changes as follows. It is 
decreased while the distance to source is increased to some
redshift $z_{opt}$ value and starts to increase behind  $z_{opt}$
up to infinite value of redshift.

In the expanding Universe the angular size of a distant object is defined according to usual equation,~\cite{wein72}:

\[
\theta = \dfrac{L}{D_A},
\label{ad}
\]

\noindent here $L$ is the size of the  object and $D_A$ is ``the angular
size distance''. This definition is in exact correspondence with
euclidean definition. But here $D_A$ does not correspond to
physical distance to the source (measured for instance, by the
light travel time). Angular size of an extragalactic
source as function of its redshift has minimum.

\subsection{The Standard $\Lambda$CDM cosmological model.}
\label{subsec:background}

The background space-time we consider in this paper corresponds to
the  Standard cosmological model with the FRW metric and
the cosmological constant $\Lambda$. The background
metric is that of the spatially flat expanding Universe,
\[
ds^2 = c^2 dt^2 -
a^2(t)d\mathbf{r}^2.
\]
The scale factor $a(t)$ is determined  by the Friedmann equation,
which can be written as follows,
\begin{equation}
\left( \frac{\dot a(t)}{a(t)} \right)^{2}=H_{0}^{2} \left[ \Omega_m \left(
\frac{a(t_0)}{a(t)} \right)^3 + \Omega_{\Lambda} \right],
\label{fried1}
\end{equation}
where $H_{0}$ is the present value of the Hubble parameter,
$a(t_0) = a_0 = 1$ is the present value of the scale factor, dot
denotes the derivative with respect to cosmic time $t$, $\Omega_m$ is matter density parameter and $\Omega_{\Lambda}$ refers to cosmological constant. 
There are two recommended set of values of global cosmological parameters \citep{WMAPreco, hin09} in Standard Cosmological Model. The first one is {\bf WMAP only} recommended values, they are: 
$H_0=71.9 \pm 2.6$ km/s/Mpc, $\Omega_m=0.26 \pm 0.03$, $\Omega_{\Lambda}=0.74 \pm 0.03$. Here we approximate recommended values in omega parameters down to two digit.

The second is {\bf WMAP + BAO + SN} recommended cosmological parameters:
$H_0=70.1 \pm 1.3$ km/s/Mpc, $\Omega_m=0.279 \pm 0.013$,
$\Omega_{\Lambda}=0.721 \pm 0.015$.

There are several  definitions of distances in Friedmannien
cosmology. They are related to usual distance definition.

\subsection{Cosmic distance in the Standard $\Lambda$CDM cosmological model.}

Cosmic distance appears as one measures the light travel time from
source to observer. The equation of light propagation obeys:

\[
ds=0,
\]

\noindent one can rewrite this equation as
\[
c dt = - a(t)dr.
\]

\noindent We choose the sign ``minus'' as corresponding to light
rays traveling from a source to an observer. Therefore, the light
ray emitted at the moment $t_e$ by the source will be detected by an
observer at the moment $t_o$ while light requests time to travel
distance $r_c$:
\[
r_c = c \int\limits_{t_e}^{t_o}\dfrac{dt}{a(t)}.
\]

Instead of cosmic distance  one have to introduce angular distance
according to definition (\ref{ad}, see below) and these distances are not equal.

\subsection{Angular size of source in the Standard $\Lambda$CDM cosmological model.}

In order to understand the  main physics of the phenomenon, we start
from the simplest case, that a source with physical size $L$ is
observed at different distances.

Angular size of the source is determined by the equation,~\cite{wein72}:

\begin{equation}
\theta = \dfrac{L}{a(t_e) r_c} ,
\label{ansz}
\end{equation}

\noindent where $a(t_e)$ is the scale factor at the moment of light
ray emission and $r_c$ is the cosmic distance to the source. 
One can introduce redshift of an epoch $t_e$ as

\[
1 + z = \dfrac{1}{a(t_e)}, 
\]

\noindent and rewrite equation for $D_A  = a(t_e) r_c$ in terms of redshift $z$:

\begin{equation}
D_A = \dfrac{c}{H_0}\dfrac{1}{1 + z} \int\limits_{0}^{z} \dfrac{d z}{\sqrt{ \Omega_m \left( 1 + z \right)^3 + \Omega_{\Lambda} }} ,
\label{angdist}
\end{equation}

Angular distance $D_A$ increases proportional to redshift for small $z \ll 1$:
\[
D_A \sim \dfrac{c z}{H_0}
\]

\noindent and start to decreases for large $z \gg 1$:

\[
D_A \sim \dfrac{2c}{H_0\sqrt{\Omega_m}} \dfrac{1}{1+z}\left( 1 - \dfrac{1}{\sqrt{1+z}}\right)
\]

\begin{figure*}
\begin{center}
\includegraphics[width=15.0cm]{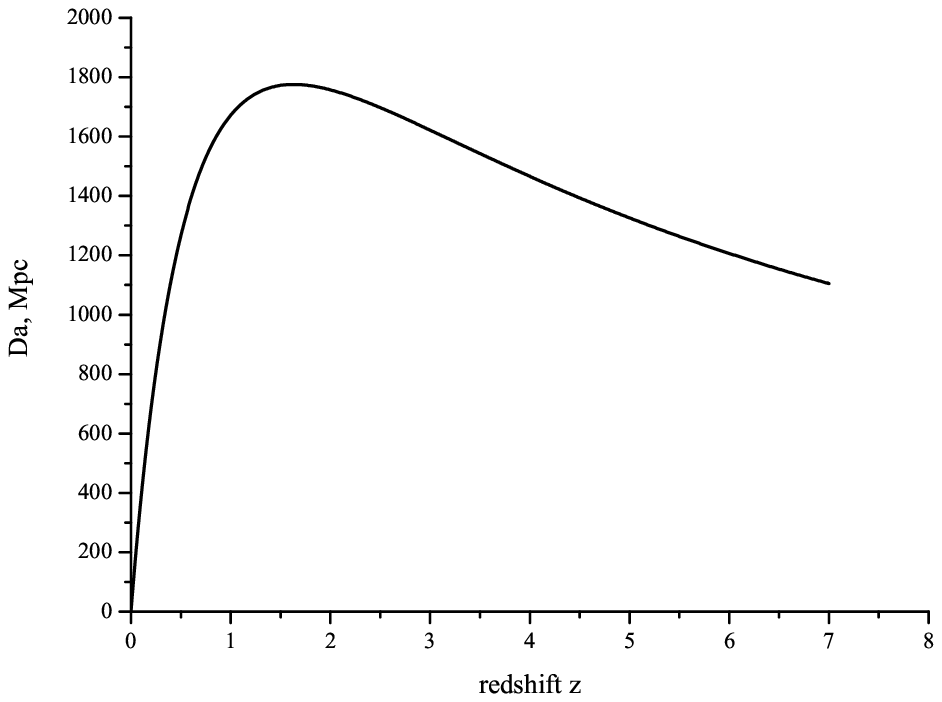}
\end{center}
\caption{The angular distance to the source (in Mpc) as function of redshift. We choose {\bf WMAP + BAO + SN} recommended cosmological parameters.}
\label{fig1}
\end{figure*}

Therefore, angular distance has maximum at some redshift. The value of redshift in maximum angular distance we designate as $z_{opt}$ and it depends on the cosmological parameters (see fig.\ref{fig1}). Apparent angular size of the source has minimum in the same redshift (fig.\ref{fig2}).

If we choose  {\bf WMAP only} recommended parameters maximal value of angular
distance is at $z=1.66$, and an object located at this distance
with physical size  of about 1 pc is subtended by an angle
$\theta=117$ $\mu as$. This is minimal angular size of an object,
while the angular size is inversly proportional to distance and it
starts to increase as $z> 1.66$. 

If we choose  {\bf WMAP + BAO + SN} recommended parameters maximal value of angular
distance is at $z=1.63$, and an object located at this distance
with physical size  of about 1 pc is subtended by an angle
$\theta=116$ $\mu as$. This is minimal angular size of an object,
while the angular size is inversly proportional to distance and it
starts to increase as $z> 1.63$. We have to mention also that {\bf WMAP + BAO + SN} variant provides us with more precise global cosmological parameters.

We have to add also a bit on the propagation of errors so that we
can assess what is desirable interval of redshift with respect to
errors in the cosmological parameters. The global cosmological parameters
like the Hubble parameter, the density parameter of matter, the density
parameter of dark energy are measured with some errors. The question
is how these errors will affect the minimal size of an extragalactic
source and how its affect available interval of redshift.

With {\bf WMAP + BAO + SN}  recommended values of the cosmological parameters and
their errors one obtains 1$\sigma$ error in $\delta \theta =3$
$\mu$as and interval of 1$\sigma$ error in redshift is $\delta z =0.005$. That is
cosmological parameter errors contribution into total uncertainty
of minimal angular size and interval of variation of the minimum
location along $z$ axes.

These estimations  show the value of the error in definition of
the minimal $ \theta $ as function of errors in the global
cosmological parameters. As we will see below, variations of
proper source size are considerably larger. Therefore the
contribution of errors of the global cosmological parameters into $\delta \theta $ and $\delta z $ intervals can be neglected and their values can be considered as exact.

\subsection{Distribution of the ICRF sources and the ``cosmological'' criterion of selection.}

\begin{figure*}
\begin{center}
\includegraphics[width=15.0cm]{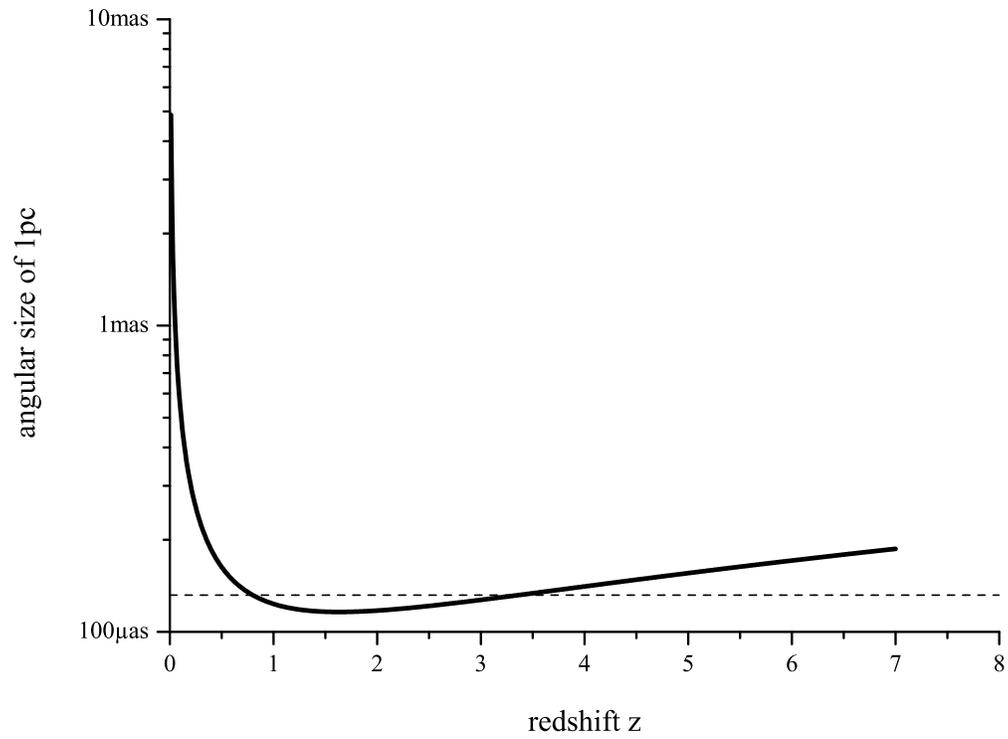}
\end{center}
\caption{The angular size of the source as function of redshift. Physical size of an object is 1 pc.  Redshift is ploted along horizontal axes.  We choose {\bf WMAP + BAO + SN} recommended cosmological parameters. Dashed line shows the interval in redshift corresponding to cosmological criterion.} \label{fig2}
\end{figure*}

Let us examine  source choice by the cosmological criterion. The
simplest way is to draw a line on the Fig.\ref{fig2} which lies
above the minimal value $\theta_{min}$ and to choose the sources
inside made well. The only question is that about the level to
draw it.

\begin{figure*}
\begin{center}
\includegraphics[width=15.0cm]{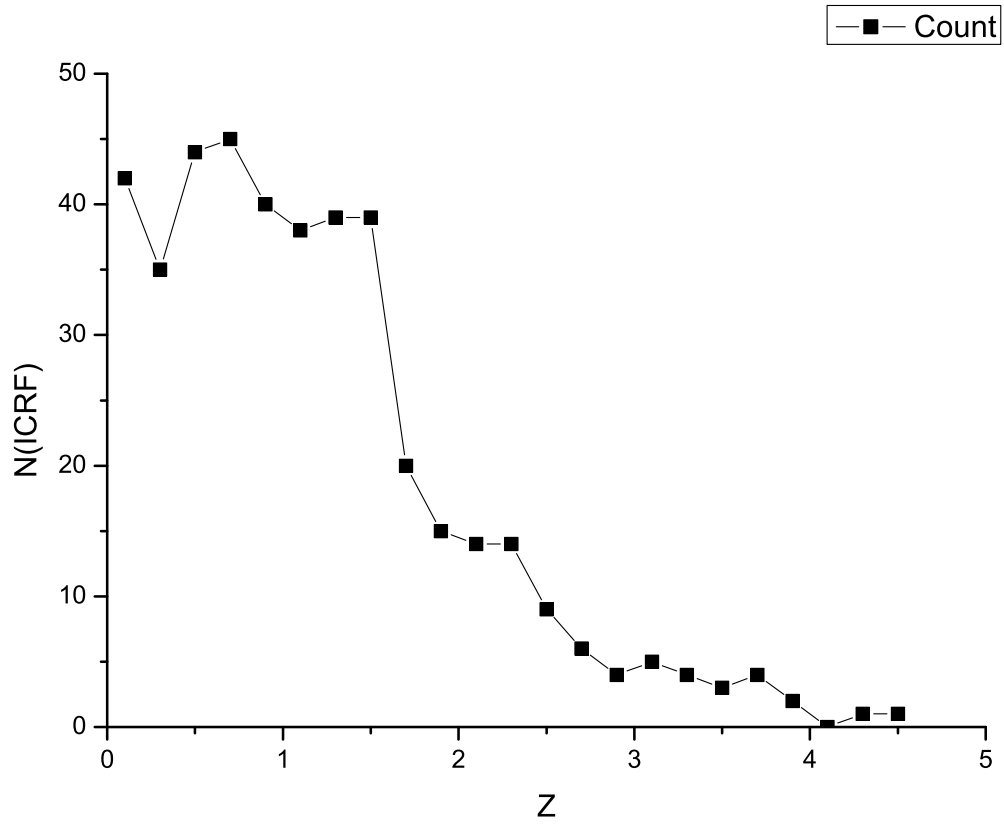}
\end{center}
\caption{The histogram of the ICRF sources distribution versus redshift.
} \label{fig3}
\end{figure*}

In order to define the level let us make a histogram  of
distribution of the ICRF sources according their redshifts. We have to mention 
that some of the ICRF sources have unknown
redshift. Therefore the  histogram (see Fig.\ref{fig3}) contains 424 sources. From the figure we conclude that for small values of 
redshifts the number of sources changes weakly. This fact is due
to the selection effect: sources at cosmologically short distances are clearly
visible and one can count and includes all of them.  

But their apparent angular motions strongly exceed apparent
angular motions of remote ones. To correct selection effect we have to take into account the
angular scale factor.

\begin{figure*}
\begin{center}
\includegraphics[width=15.0cm]{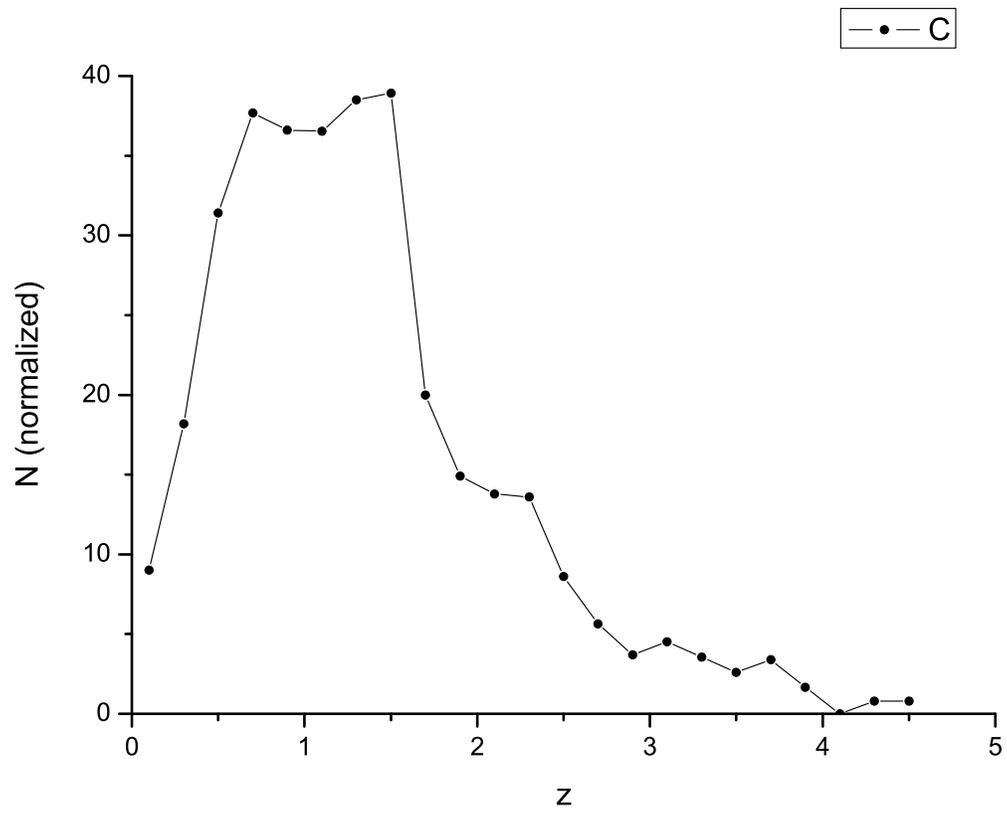}
\end{center}
\caption{The weighted distribution of the ICRF sources.} \label{fig4}
\end{figure*}

To take it into account one can draw the weighted histogram which is 
the production $N(z) \cdot D_A$. The ``weight'' is angular distance from
an observer to a source $D_A$. Using this weighted
histogram one can immediately see the best value of the redshift
to make selection (see Fig.\ref{fig4}). If redshift is less than
$z=0.8$ then the normalized quantity $N(z) \cdot D_A$ decreases sharply. 
Therefore the value $z=0.8$ can be selected as the low boundary of the interval.

Note that for $z=0.8$ the growth  of an apparent angular size of a
standard object is $30\%$. The corresponded horizontal line
drawing in Fig.\ref{fig2} intersects the curve $\theta(z)$ at $z
\approx 3.0$.

The value $z\approx3.0$ is important in the Universe history. This value corresponds 
approximately to the re-ionization of He II and change
intergalactic medium and conditions of quasars observations, their
spectra and other properties of these objects, \cite{dor09, Theuns2002}.

Therefore as we have seen it looks like appropriate to draw the line
on the Fig.\ref{fig2} intersecting the curve $\theta(z)$ in
points $z_s = 0.8$ and $z_f = 3.0$. Restricting the cosmological
criterion by the ICRF sources possessing redhifts from the
interval $0.8 \le z \le 3.0$, we receive a list of 239 sources.

\section{The ICRF source model}\label{psize}

\begin{figure*}
\begin{center}
\includegraphics[width=15.0cm]{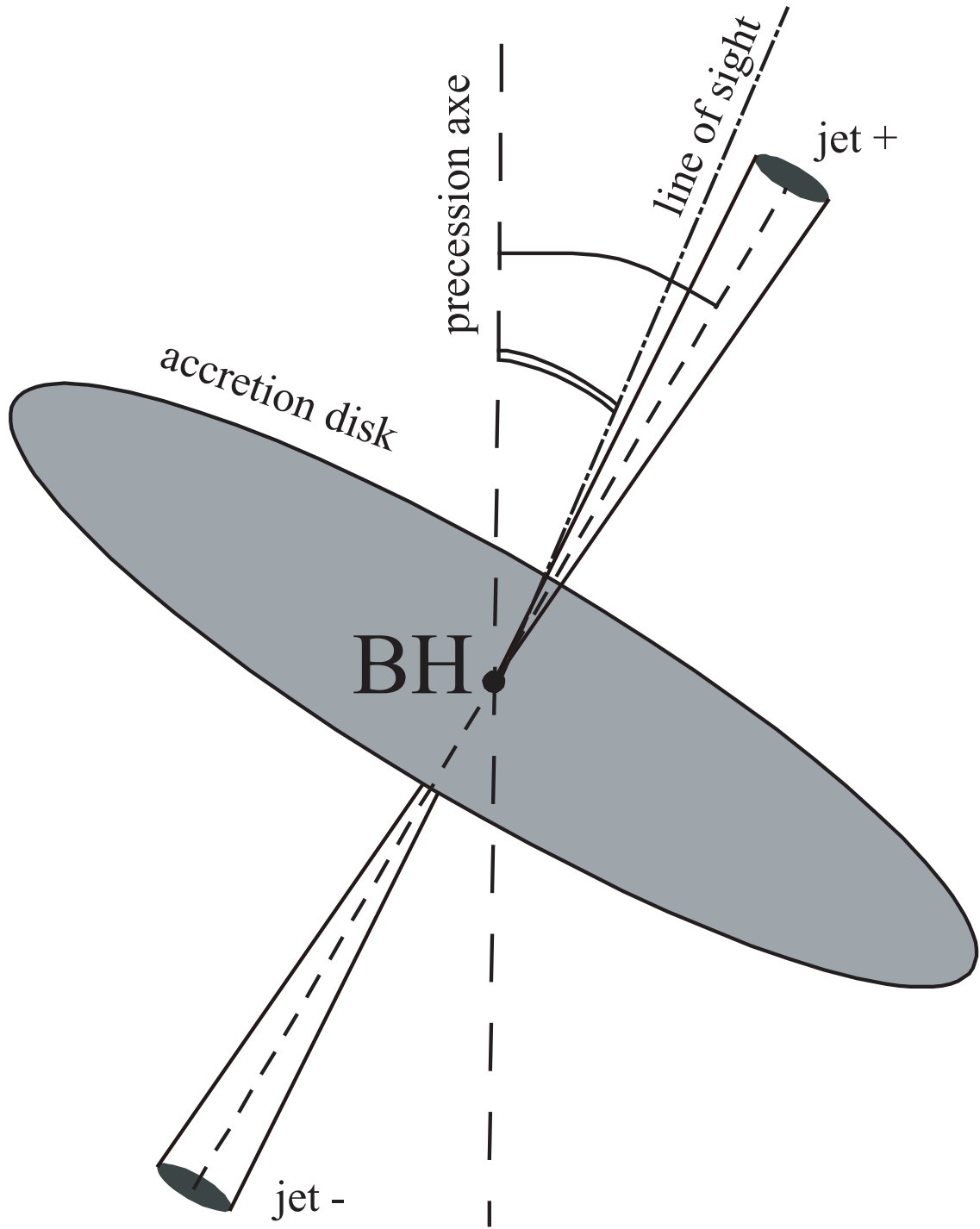}
\end{center}
\caption{This figure represents the Blandford--Rees model. Central black hole (BH) is surrounded by accretion disk, and two jets from polar regions. The ``jet+'' is directed to observer. The small black ellipse ending the ``+'' cone represents ``hot spot''.} 
\label{fig5}
\end{figure*}

Let us discuss the ICRF sources model (Blandford-Rees model or BR model below) which
we chosen as basic. According this model source is a system of a supermassive black
hole ($\sim10^6\div10^{10}\;M_{\sun}$), surrounded by an  accretion disk  
with two opposite jets from polar regions (see, Fig.~\ref{fig5}).
Jets are main sources of radio emission, \cite{bla79, beg84}.
It is worth noting that there are evidences of compact radio
source existing. Its location coincides with a black hole position,
\cite{jack06}. If some of the ICRF sources are compact and possess the
radiation from black hole horizon or coronal regions then these
sources are almost stationary, without any apparent motions. They satisfy complitely the kinematical principle,~\cite{Kov} and therefore these sources holding the constant coordinates are the best set for astrometric tools while other radio sources with inertial
apparent motions (with velocities close to the speed of light) are
``hot spots'' in jets.

In the paper \cite{zhar09} the BR model has been considered with regard
to the ICRF sources. Quiet radioquasar jet was treated as
stationary structure. Distribution of brightness over jet was stationary.
The maximal brightness region was identified with the ``hot spot''
which was considered as a radio source. Position of the spot was
associated with the ICRF source position on the celestial sphere.

Note  once more that if jet is stationary or the ICRF source is
compact one and lies near black hole horizon, then this ICRF
source is treated as stationary and can be considered as 
the ``standard ICRF source''.

However  jet can precess that leads to source motion. Jet
represents the high temperature plasma moving with very high
velocity comparable with the speed of light. If the motion has small angle
in an observer direction, the source apparent motion can be faster
than the speed of light. The simplest model of stationary jet motion is
motion due precession. For the realistic periods of jet precession $P
\sim 1000 \div 10 000$ years the velocity of apparent motion
of the ICRF sources is $10 \div 20$ $\mu$as/yr.

As far as precession periods of the sources are significantly
larger than time of observation, their motions can be treated
as linear and predictable with high accuracy for time interval
of $10 \div 30$ years. 

Here we have to mention that precession velocity $\Omega$
for relativistic jets is almost constant. This experimental fact 
comes from the observation of binaries in the Milky Way. The black hole system in the
object SS433 (some astronomers call it as microquasar) was observed during interval of 30 years. This system is in good agreement with BR model: a black hole of stellar mass surrounded with accretion disk and two jets from polar regions. Jets are precesing, the observed precession period is 162 days that is significantly less than a supermassive black hole precession period. So, quasar differs from SS433 in mass of black hole and precession period only. Other properties are similar to BR model. 

The precession period stability is $5\cdot 10^{-5}$ and the
``glitches'' do not exceed 6 \% \cite{cher08}. This way
one can expect that the supermassive black hole precession
stability would be at the same order of magnitude and therefore the
predictability of jet motions in the ICRF sources is as high as 6\%.

In the papers ~\cite{zach99,  net02, ass03, ass07} the catalogue of
optical positions of 172 ICRF sources was composed. In the first
paper it was already appeared the significant spacial difference
between optical and radio quasar components. The error in
coordinates defining of chosen source was approximately 50 mas, as
an average shift in source groups was changed from 80 to 90 mas;
error of mean value was about 6 mas.

The errors consist less than 10\% from the average shift. From our
point of view the significant difference between optical and radio
source positions indicates that the BR model is
appropriate for our purposes.

Taking into account all reasons we assume that the
position difference does not due systematic errors. It is real
physical effect.

\begin{figure*}
\begin{center}
\includegraphics[width=15.0cm]{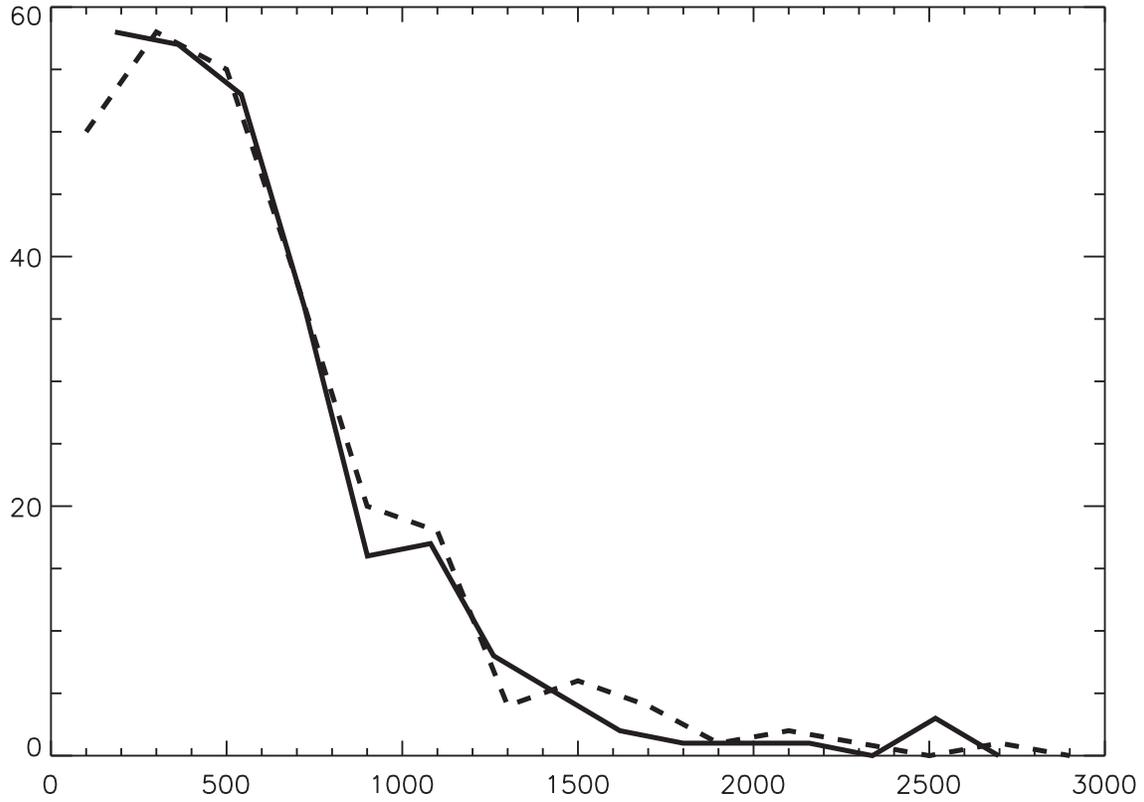}
\end{center}
\caption{
The observed distribution of projection distance between optical and radio component in the ICRF sample is plotted. The observed distribution is plotted as a dashed line. The solid line represents simulated distribution. The last is formed by the multiplication of two random functions: gamma distribution and sinus of uniformly distributed angle.
} 
\label{fig6}
\end{figure*}

We used the list of angular distances between optical and radio
components from the papers listed above to estimate the physical distances
from jet beginning to ``hot spot'' and to calculate average values of the distribution of the physical distances. One can draw the distribution of physical distances obtained from the observational data (see Fig.~\ref{fig6}). As far as the sources number
is not very large, the systematic errors due to small sampling are
big enough, but they are convenient for our purpose to define the
type of distribution function (it is similar to gamma distribution function). 

Propagation of the corresponding errors provides us with error of individual 
measurements of the order of 16 $\mu$as.

Recently we have discussed (see \cite{zhar09}) the apparent motion
of astrometric radio sources. Taking into account the fact that
many of the ICRF sources demonstrate apparent superluminal motion, we
have concluded that such sources are plasma clouds or jets. We
also have estimated the consequent precession periods. For
sampling of 6 sources the period belongs to interval from thousand to
several thousand years.

It is easy to calculate the apparent motion of ``hot spot'' in jet
of found size and precession period $T$. The velocity is about
$\sim 2\pi R/T$. For periods of the order of thousand or several
thousand years the velocity is close to the speed of light or even
exceed it. 

The angular velocity of source motion can be also easy estimated.
For a source belonging to the redshift interval $z \sim 0.8 \div
3.0$ one can recalculate physical distance to angular one as
follows:
\[
s \approx 7.5 kpc/1'',
\]
\noindent which for the object size of 500 pc gives angular size
$\sim$ 60 mas. For the precession period from the interval $1000 \div
10000$ years the corresponding angular velocities will be $6\div
60$ $\mu$as/yr. This result is in accordance with estimations of
apparent motions of the ICRF sources done in \cite{mcm03, mcmma07,
tit08, tit09}.

Now we will discuss one more criterion for source selection,
``kinematical'' one.

In the frames of the BR model we consider several types of
source motions. The first one is the precession motion of jet. The
second one is acceleration of clouds of interstellar medium, which
penetrate into jet time to time. The radio source brightness center
begin to move with acceleration. Ordinary time intervals when
cloud acceleration occurs is from three to ten years. During this
time interval sources demonstrate accelerated motion.

This kind of radio sources we consider as ``bad'' ones because of
the following reason. The process of cloud penetration into jet
and its acceleration is stochastic process and its beginning is
almost unpredictable. 

The sources which are stationary ones during all observational
time or having only linear uniform motion we consider as ``good''.
Characteristics of such motions will not change during long time
interval.

Therefore we have to choose into the ICRF list only sources with
uniform and linear motion. This criterion we call ``kinematical''.

During observational time period ($\sim$ 30) of the ICRF sources ``hot
spot'' will pass negligibly small part of its whole path and
therefore for an observer it will look like uniform and linear
motion.

We used code ARIADNA elaborated by \citep{zhar09a} to analyze the VLBI data. As the result of analysis we got time series and polynomial approximation of source coordinates as function of time:
\begin{equation}
\alpha(t) = \alpha_0 + \alpha_1 t + \frac{1}{2} \alpha_2 t
^2 + \frac{1}{3!} \alpha_3 t^3 +...  ,
\label{alfa}
\end{equation}

\begin{equation}
\delta(t) = \delta_0 + \delta_1 t + \frac{1}{2} \delta_2 t
^2 + \frac{1}{3!} \delta_3 t^3 +... ,
\label{delta}
\end{equation}

\noindent here $t$ is the years from epoch J2000.0

We select sources which time series have only zero and first power significant coefficients and call them as ``good'' candidates. We eliminate sources with quadratic and cubic significant coefficients and call them as ``bad'' candidates. 

After selection of sources as ``bad'' and ``good'' according the ``kinematical'' criterion we obtained the final list of 137 sources (Table \ref{table-proposal}). Zero power coefficients in time series of polynomial approximation were used for calculation of correction to the ICRF coordinates at J2000.0, and linear coefficients were used for calculation of proper motion of the sources. The table contains 12 columns. The first column is number of the ICRF source. The second one is right ascension at J2000.0, the third column is declination of the source at J2000.0. The 4th column is redshift of the source and the 5th column is type of the source. QSO is quasar type sources, BLL is BL Lacertae type sources, AGN is Active Galactic Nuclei type sources. In the case of AGN type sources we check optical image of the source and select only point-like sources. The 6th and the 7th columns are errors of right ascension and declination in milliarcseconds, the 8th and 9th columns are proper motion along right ascension and declination respectively in observer's plane in units mas/year, and the last (10th) column is flag of motion. If this flag exists, the source is moving otherwise the source is fixed. Flag 2 means that source is moving at 2$\sigma$ confidence level. Flag 3 means that source is moving at 3$\sigma$ confidence level.

We compared our list of ``good'' sources with ``good'' source list of \cite{fei06}. Because  author uses statistical and astronomical criteria and we used physical criteria (cosmological and kinematical) we have to compare only names of candidates to obtain the cross-list. 

List of \cite{fei06} contains 362 sources which are selected in a two-step process, as follows:

\begin{enumerate}
\item A first selection is made on the basis of continuity criteria for one year
weighted average coordinates.

\begin{itemize}
\item[a.] Length of observation period longer than five years.
\item[b.] Not less than two observations of the source in a given session.
\item[c.] One-year average coordinates based on at least three observations.
\item[d.] Not more than three successive years with no observations, conditions (b) and (c) being met.
\item[e.] At least half of the one-year averages available over the source observation time span.
\end{itemize}

This first screening keeps 362 sources for the years centered at
1990.0 through 2002.0. These include 141 defining sources, 130
candidates, 87 other and 4 new sources, i.e. 67\% of the defining
sources, 44\% of the candidates and 85\% of the others.

\item The time series of yearly values of   $\Delta \alpha \cos\delta$ and  $\Delta \delta$ are then analysed in order to derive

\begin{itemize}
\item[a.] the linear drift (least squares estimation) and the normalized drift. The normalized linear drift is the absolute value of the least-square derived linear drift divided by its formal uncertainty; 
\item[b.] the Allan standard deviation for a one-year sampling time.
\end{itemize}
\end{enumerate}

Among these 362 sources author selected sources which satisfy the following conditions: Allan Standard deviation is less than 200 $\mu$as and linear drift is less than 50 $\mu$as/year. 199 sources which satisfy these conditions are considered as stable. 

Intersection of our list with primary list of the 362 ICRF2 sources is 120, while the intersection with subset of ''stable'' sources is only 70. Other 50 sources of our list reveal large apparent motion $\le$ 50 $\mu$as/year (38 sources), and 12 sources reveal large Allan Standard deviation $\le$ 200 $\mu$as.

In spite of the fact that the intersection of our list with ''stable'' list of \cite{fei06} is only 70, one have to compare only part of our list (120 - 38 = 82) because 38 can not appear in the ''stable'' list due to selection criteria. After this adjustment the intersection of two lists became more impressive.

Though the methods used by \cite{fei06} are based on different principles of selection, the high intersection of the source lists indicates the quality of our criteria. In other words the ``cosmological'' and ``kinematical'' criteria introduced by our group work very productive to source selection and shed new light on the nature of stability of the ICRF sources.

\section{Conclusions}

The main conclusion concerning the selection of new set of the ICRF sources is that it is necessary to use astronomical and statistical criteria. 
The astronomical criteria mean sources to be bright, 
uniformly distributed into the sky sphere and so on.

In the present paper we considered several propositions for additional principles to formation of new ICRF catalogue. As source selection criteria we discussed two ones, ``cosmological'' and ``kinematical''. Their physical meaning is clear. It is based on the assumption that the apparent motions of quasars (for angular shifts at hundred $\mu$as) are related with real motion inside these radio sources. Therefore angular shift corresponds to some physical shift of ``hot spot'' inside radio source. It leads to two simple criteria. The first one is that we have to select sources at redshifts which corresponds to minimal angular shift. The second one is that we have to choose well known and predictable motions for long time interval. From these two criteria we found the optimal redshif interval $0.8 \le z \le 3.0$, at that we have to choose sources. We also chosen the linear motion as the most simple and predictable physical motion.

Certainly these two criteria are no dogma. So, in the first article, \cite{zhar09}, we found several sources, for example, $1150+497$ (``defining''), $1738+476$ (``candidate''), which do not belong to interval $0.8 \le z \le 3.0$ but demonstrate the high quality properties by statistical selection criterion. It may be related with small precession velocity. It leads to ``hot spot'' way to be significantly less than the corresponding way of source being even in ``good'' redshift interval. Without any doubt these sources have to be included in future catalogues. In this sense we can call our criteria as ``sufficient'' but not ``necessary''.

In addition one has to include into the principles of catalogue formation not only source coordinates but also their linear velocities. The modern accuracy of coordinate observations by the VLBI is so high that a part of radiosources has apparent proper motions, which are not related with motion of central massive object, and are related with ``hot spot'' motion of the radiosource, which produces radio emission.

\section*{Acknowledgemets} 
This work has been partially supported by Russian Foundation for
Basic Research grants 07-02-01034a (M.S., V.S., and O.S. ) and 08-02-00971à (V.Z.), grant of the President of RF MK-2503.2008.2 (O.S.).

\newpage

\clearpage

\newpage

\begin{table*}
\begin{minipage}{150mm}
\caption{The ICRF2 source list selected by cosmological and kinematical criteria}
\label{table-proposal}
\begin{tabular}{@{}lrrccrrrrl}
\hline 
ICRF     & \multicolumn{1}{c}{$\alpha_{2000}$} & \multicolumn{1}{c}{$\delta_{2000}$}&   Z&   src  & $\varepsilon\Delta\alpha$ & $\varepsilon\Delta\delta$ & \multicolumn{1}{c}{$\mu_{\alpha}$}    & \multicolumn{1}{c}{$\mu_{\delta}$}    & flag\\
\hline 
$0013-005$ & 00 \ 16 \ 11.088556 & $-$00\ 15\ 12.44537 & 1.57 &  QSO & 0.22 & 0.32 &  0.002$\pm$0.096 &  0.041$\pm$0.133 &       \\
$0016+731$ & 00 \ 19 \ 45.786430 & $+$73\ 27\ 30.01750 & 1.78 &  QSO & 0.07 & 0.06 & -0.009$\pm$0.010 & -0.012$\pm$0.009 &       \\
$0035+413$ & 00 \ 38 \ 24.843614 & $+$41\ 37\ 06.00067 & 1.35 &  QSO & 0.16 & 0.24 & -0.049$\pm$0.028 & -0.045$\pm$0.044 &       \\
$0106+013$ & 01 \ 08 \ 38.771076 & $+$01\ 35\ 00.31713 & 2.11 &  QSO & 0.02 & 0.03 &  0.042$\pm$0.004 &  0.009$\pm$0.006 & 3     \\
$0119+041$ & 01 \ 21 \ 56.861698 & $+$04\ 22\ 24.73434 & 0.64 &  QSO & 0.03 & 0.04 &  0.000$\pm$0.005 &  0.003$\pm$0.006 &       \\
$0133+476$ & 01 \ 36 \ 58.594810 & $+$47\ 51\ 29.10004 & 0.86 &  QSO & 0.01 & 0.02 & -0.004$\pm$0.003 & -0.009$\pm$0.004 & 2     \\
$0146+056$ & 01 \ 49 \ 22.370914 & $+$05\ 55\ 53.56855 & 2.35 &  QSO & 0.15 & 0.21 &  0.061$\pm$0.057 &  0.130$\pm$0.081 &       \\
$0149+218$ & 01 \ 52 \ 18.059047 & $+$22\ 07\ 07.70002 & 1.32 &  QSO & 0.10 & 0.17 & -0.024$\pm$0.021 & -0.040$\pm$0.032 &       \\
$0151+474$ & 01 \ 54 \ 56.289911 & $+$47\ 43\ 26.53911 & 1.03 &  QSO & 0.12 & 0.19 &  0.024$\pm$0.021 &  0.043$\pm$0.033 &       \\
$0208-512$ & 02 \ 10 \ 46.200415 & $-$51\ 01\ 01.89190 & 1.00 &  BLL & 0.08 & 0.10 & -0.012$\pm$0.013 & -0.012$\pm$0.016 &       \\
$0212+735$ & 02 \ 17 \ 30.813380 & $+$73\ 49\ 32.62177 & 2.37 &  QSO & 0.06 & 0.05 &  0.051$\pm$0.006 & -0.014$\pm$0.005 & 3     \\
$0215+015$ & 02 \ 17 \ 48.954740 & $+$01\ 44\ 49.69899 & 1.72 &  QSO & 0.10 & 0.19 & -0.023$\pm$0.033 & -0.068$\pm$0.061 &       \\
$0248+430$ & 02 \ 51 \ 34.536776 & $+$43\ 15\ 15.82862 & 1.31 &  QSO & 0.09 & 0.14 & -0.020$\pm$0.016 &  0.082$\pm$0.026 & 3     \\
$0256+075$ & 02 \ 59 \ 27.076632 & $+$07\ 47\ 39.64321 & 0.89 &  QSO & 0.09 & 0.25 & -0.011$\pm$0.019 &  0.037$\pm$0.053 &       \\
$0306+102$ & 03 \ 09 \ 03.623519 & $+$10\ 29\ 16.34083 & 0.86 &  QSO & 0.08 & 0.12 &  0.019$\pm$0.025 & -0.006$\pm$0.036 &       \\
$0319+121$ & 03 \ 21 \ 53.103501 & $+$12\ 21\ 13.95376 & 2.67 &  QSO & 0.14 & 0.21 & -0.008$\pm$0.028 & -0.032$\pm$0.042 &       \\
$0400+258$ & 04 \ 03 \ 05.586051 & $+$26\ 00\ 01.50278 & 2.11 &  QSO & 0.14 & 0.15 &  0.039$\pm$0.024 & -0.042$\pm$0.026 &       \\
$0402-362$ & 04 \ 03 \ 53.749902 & $-$36\ 05\ 01.91300 & 1.42 &  QSO & 0.06 & 0.08 & -0.005$\pm$0.013 & -0.013$\pm$0.017 &       \\
$0405-385$ & 04 \ 06 \ 59.035302 & $-$38\ 26\ 28.04093 & 1.29 &  QSO & 0.25 & 0.19 &  0.018$\pm$0.046 &  0.060$\pm$0.039 &       \\
$0420-014$ & 04 \ 23 \ 15.800727 & $-$01\ 20\ 33.06530 & 0.92 &  QSO & 0.06 & 0.08 &  0.014$\pm$0.008 & -0.016$\pm$0.009 &       \\
$0434-188$ & 04 \ 37 \ 01.482725 & $-$18\ 44\ 48.61344 & 2.70 &  QSO & 0.11 & 0.17 & -0.004$\pm$0.020 &  0.017$\pm$0.032 &       \\
$0454+844$ & 05 \ 08 \ 42.363508 & $+$84\ 32\ 04.54398 & 1.34 &  BLL & 0.08 & 0.08 & -0.020$\pm$0.015 &  0.039$\pm$0.016 & 2     \\
$0454-234$ & 04 \ 57 \ 03.179224 & $-$23\ 24\ 52.01994 & 1.00 &  QSO & 0.02 & 0.02 &  0.006$\pm$0.004 & -0.026$\pm$0.005 & 3     \\
$0457+024$ & 04 \ 59 \ 52.050666 & $+$02\ 29\ 31.17634 & 2.38 &  QSO & 0.11 & 0.14 &  0.014$\pm$0.022 &  0.030$\pm$0.028 &       \\
$0458-020$ & 05 \ 01 \ 12.809887 & $-$01\ 59\ 14.25623 & 2.29 &  QSO & 0.01 & 0.02 & -0.014$\pm$0.003 & -0.002$\pm$0.004 & 3     \\
$0528+134$ & 05 \ 30 \ 56.416744 & $+$13\ 31\ 55.14952 & 2.07 &  QSO & 0.01 & 0.01 &  0.003$\pm$0.002 &  0.004$\pm$0.002 &       \\
$0537-441$ & 05 \ 38 \ 50.361541 & $-$44\ 05\ 08.93896 & 0.89 &  QSO & 0.05 & 0.06 &  0.001$\pm$0.008 & -0.016$\pm$0.010 &       \\
$0552+398$ & 05 \ 55 \ 30.805609 & $+$39\ 48\ 49.16498 & 2.36 &  QSO & 0.01 & 0.01 &  0.004$\pm$0.001 & -0.002$\pm$0.001 & 3     \\
$0602+673$ & 06 \ 07 \ 52.671670 & $+$67\ 20\ 55.40988 & 1.97 &  QSO & 0.04 & 0.04 & -0.003$\pm$0.008 & -0.022$\pm$0.008 & 2     \\
$0605-085$ & 06 \ 07 \ 59.699234 & $-$08\ 34\ 49.97807 & 0.87 &  QSO & 0.22 & 0.24 & -0.004$\pm$0.046 & -0.038$\pm$0.047 &       \\
$0609+607$ & 06 \ 14 \ 23.866188 & $+$60\ 46\ 21.75544 & 2.70 &  QSO & 0.12 & 0.12 & -0.013$\pm$0.020 &  0.032$\pm$0.020 &       \\
$0615+820$ & 06 \ 26 \ 03.006179 & $+$82\ 02\ 25.56768 & 0.71 &  QSO & 0.09 & 0.16 &  0.022$\pm$0.018 & -0.027$\pm$0.029 &       \\
$0727-115$ & 07 \ 30 \ 19.112469 & $-$11\ 41\ 12.60045 & 1.59 &  QSO & 0.01 & 0.02 &  0.022$\pm$0.002 & -0.007$\pm$0.003 & 3     \\
$0743+259$ & 07 \ 46 \ 25.874166 & $+$25\ 49\ 02.13486 & 2.98 &  QSO & 0.06 & 0.10 &  0.024$\pm$0.011 & -0.018$\pm$0.017 & 2     \\
$0748+126$ & 07 \ 50 \ 52.045731 & $+$12\ 31\ 04.82817 & 0.89 &  QSO & 0.10 & 0.13 &  0.027$\pm$0.023 &  0.043$\pm$0.032 &       \\
$0804+499$ & 08 \ 08 \ 39.666275 & $+$49\ 50\ 36.53045 & 1.43 &  QSO & 0.01 & 0.01 &  0.005$\pm$0.002 & -0.008$\pm$0.003 & 2     \\
$0805+410$ & 08 \ 08 \ 56.652039 & $+$40\ 52\ 44.88888 & 1.42 &  QSO & 0.02 & 0.03 & -0.004$\pm$0.005 & -0.009$\pm$0.006 &       \\
$0808+019$ & 08 \ 11 \ 26.707319 & $+$01\ 46\ 52.21999 & 1.15 &  BLL & 0.07 & 0.09 &  0.027$\pm$0.017 &  0.033$\pm$0.023 &       \\
$0812+367$ & 08 \ 15 \ 25.944829 & $+$36\ 35\ 15.14842 & 1.03 &  QSO & 0.18 & 0.24 &  0.008$\pm$0.032 &  0.011$\pm$0.041 &       \\
$0828+493$ & 08 \ 32 \ 23.216695 & $+$49\ 13\ 21.03823 & 1.26 &  BLL & 0.22 & 0.20 &  0.024$\pm$0.045 & -0.028$\pm$0.039 &       \\
$0833+585$ & 08 \ 37 \ 22.409723 & $+$58\ 25\ 01.84515 & 2.10 &  QSO & 0.12 & 0.12 & -0.048$\pm$0.022 &  0.025$\pm$0.021 & 2     \\
$0836+710$ & 08 \ 41 \ 24.365247 & $+$70\ 53\ 42.17323 & 2.22 &  QSO & 0.10 & 0.12 & -0.020$\pm$0.018 & -0.010$\pm$0.022 &       \\
$0839+187$ & 08 \ 42 \ 05.094181 & $+$18\ 35\ 40.99055 & 1.27 &  QSO & 0.10 & 0.23 &  0.012$\pm$0.027 &  0.009$\pm$0.062 &       \\
$0850+581$ & 08 \ 54 \ 41.996390 & $+$57\ 57\ 29.93927 & 1.32 &  QSO & 0.13 & 0.17 & -0.005$\pm$0.024 & -0.054$\pm$0.032 &       \\
$0917+449$ & 09 \ 20 \ 58.458483 & $+$44\ 41\ 53.98504 & 2.18 &  QSO & 0.11 & 0.15 & -0.013$\pm$0.019 & -0.043$\pm$0.028 &       \\
$0923+392$ & 09 \ 27 \ 03.013911 & $+$39\ 02\ 20.85193 & 0.70 &  QSO & 0.01 & 0.01 &  0.032$\pm$0.001 & -0.012$\pm$0.002 & 3     \\
$0945+408$ & 09 \ 48 \ 55.338149 & $+$40\ 39\ 44.58711 & 1.25 &  QSO & 0.11 & 0.16 &  0.004$\pm$0.021 & -0.031$\pm$0.029 &       \\
$0952+179$ & 09 \ 54 \ 56.823622 & $+$17\ 43\ 31.22232 & 1.48 &  QSO & 0.08 & 0.11 &  0.014$\pm$0.018 &  0.024$\pm$0.026 &       \\
$0953+254$ & 09 \ 56 \ 49.875361 & $+$25\ 15\ 16.04966 & 0.71 &  QSO & 0.02 & 0.03 &  0.051$\pm$0.005 &  0.035$\pm$0.006 & 3     \\
$0955+476$ & 09 \ 58 \ 19.671647 & $+$47\ 25\ 07.84248 & 1.87 &  QSO & 0.01 & 0.01 & -0.018$\pm$0.002 & -0.003$\pm$0.002 & 3     \\
$1011+250$ & 10 \ 13 \ 53.428737 & $+$24\ 49\ 16.44113 & 1.64 &  QSO & 0.14 & 0.18 & -0.030$\pm$0.029 &  0.024$\pm$0.034 &       \\
$1014+615$ & 10 \ 17 \ 25.887533 & $+$61\ 16\ 27.49678 & 2.80 &  QSO & 0.17 & 0.12 & -0.012$\pm$0.065 & -0.086$\pm$0.042 & 2     \\
$1020+400$ & 10 \ 23 \ 11.565626 & $+$39\ 48\ 15.38537 & 1.25 &  QSO & 0.20 & 0.26 &  0.039$\pm$0.038 &  0.012$\pm$0.043 &       \\
$1022+194$ & 10 \ 24 \ 44.809594 & $+$19\ 12\ 20.41524 & 0.83 &  QSO & 0.07 & 0.12 &  0.004$\pm$0.020 &  0.042$\pm$0.032 &       \\
$1030+415$ & 10 \ 33 \ 03.707851 & $+$41\ 16\ 06.23295 & 1.12 &  QSO & 0.11 & 0.17 & -0.066$\pm$0.021 &  0.038$\pm$0.031 & 3     \\
$1032-199$ & 10 \ 35 \ 02.155272 & $-$20\ 11\ 34.35975 & 2.20 &  QSO & 0.26 & 0.53 &  0.122$\pm$0.052 &  0.036$\pm$0.104 & 2     \\
$1038+064$ & 10 \ 41 \ 17.162502 & $+$06\ 10\ 16.92372 & 1.27 &  QSO & 0.08 & 0.15 & -0.009$\pm$0.020 &  0.122$\pm$0.037 & 3     \\
$1038+528$ & 10 \ 41 \ 46.781638 & $+$52\ 33\ 28.23129 & 0.68 &  QSO & 0.05 & 0.06 & -0.003$\pm$0.012 &  0.014$\pm$0.015 &       \\
$1039+811$ & 10 \ 44 \ 23.062564 & $+$80\ 54\ 39.44302 & 1.26 &  QSO & 0.06 & 0.06 & -0.022$\pm$0.011 & -0.007$\pm$0.010 & 2     \\
$1044+719$ & 10 \ 48 \ 27.619896 & $+$71\ 43\ 35.93845 & 1.15 &  QSO & 0.01 & 0.01 & -0.008$\pm$0.003 &  0.018$\pm$0.003 & 3     \\
$1053+704$ & 10 \ 56 \ 53.617497 & $+$70\ 11\ 45.91583 & 2.49 &  QSO & 0.08 & 0.08 & -0.023$\pm$0.018 & -0.045$\pm$0.018 & 2     \\
$1104-445$ & 11 \ 07 \ 08.694136 & $-$44\ 49\ 07.61834 & 1.60 &  QSO & 0.41 & 0.30 & -0.078$\pm$0.074 &  0.052$\pm$0.056 &       \\
$1111+149$ & 11 \ 13 \ 58.695094 & $+$14\ 42\ 26.95260 & 0.87 &  QSO & 0.16 & 0.23 & -0.004$\pm$0.029 & -0.030$\pm$0.042 &       \\
\hline                        
\end{tabular} 
\end{minipage}
\end{table*}

\addtocounter{table}{-1}
\begin{table*}
\begin{minipage}{150mm}
\caption{continued}
\begin{tabular}{@{}lrrccrrrrl}
\hline 
ICRF     & \multicolumn{1}{c}{$\alpha_{2000}$} & \multicolumn{1}{c}{$\delta_{2000}$}&   Z&   src  & $\varepsilon\Delta\alpha$ & $\varepsilon\Delta\delta$ & \multicolumn{1}{c}{$\mu\alpha$}    & \multicolumn{1}{c}{$\mu\delta$}    & flag\\
\hline 
$1116+128$ & 11 \ 18 \ 57.301440 & $+$12\ 34\ 41.71811 & 2.12 &  QSO & 0.12 & 0.21 & -0.001$\pm$0.023 & -0.055$\pm$0.042 &       \\
$1124-186$ & 11 \ 27 \ 04.392430 & $-$18\ 57\ 17.44157 & 1.05 &  QSO & 0.04 & 0.04 &  0.022$\pm$0.008 & -0.031$\pm$0.008 & 3     \\
$1130+009$ & 11 \ 33 \ 20.055798 & $+$00\ 40\ 52.83717 & 1.63 &  QSO & 0.23 & 0.33 & -0.049$\pm$0.040 &  0.035$\pm$0.058 &       \\
$1144+402$ & 11 \ 46 \ 58.297906 & $+$39\ 58\ 34.30458 & 1.09 &  QSO & 0.06 & 0.08 & -0.003$\pm$0.008 &  0.013$\pm$0.010 &       \\
$1144-379$ & 11 \ 47 \ 01.370688 & $-$38\ 12\ 11.02350 & 1.05 &  QSO & 0.06 & 0.07 &  0.036$\pm$0.011 & -0.032$\pm$0.012 & 3     \\
$1156+295$ & 11 \ 59 \ 31.833913 & $+$29\ 14\ 43.82691 & 0.73 &  QSO & 0.01 & 0.02 & -0.004$\pm$0.003 &  0.002$\pm$0.004 &       \\
$1216+487$ & 12 \ 19 \ 06.414733 & $+$48\ 29\ 56.16495 & 1.08 &  QSO & 0.12 & 0.13 &  0.029$\pm$0.021 &  0.018$\pm$0.022 &       \\
$1219+044$ & 12 \ 22 \ 22.549618 & $+$04\ 13\ 15.77625 & 0.97 &  QSO & 0.01 & 0.02 &  0.012$\pm$0.004 &  0.016$\pm$0.005 & 3     \\
$1237-101$ & 12 \ 39 \ 43.061435 & $-$10\ 23\ 28.69263 & 0.75 &  QSO & 0.12 & 0.20 & -0.054$\pm$0.032 &  0.016$\pm$0.044 &       \\
$1244-255$ & 12 \ 46 \ 46.802040 & $-$25\ 47\ 49.28878 & 0.64 &  QSO & 0.13 & 0.12 &  0.020$\pm$0.025 & -0.050$\pm$0.023 & 2     \\
$1252+119$ & 12 \ 54 \ 38.255603 & $+$11\ 41\ 05.89509 & 0.87 &  QSO & 0.16 & 0.16 & -0.044$\pm$0.029 &  0.044$\pm$0.030 &       \\
$1255-316$ & 12 \ 57 \ 59.060776 & $-$31\ 55\ 16.85185 & 1.92 &  QSO & 0.13 & 0.19 & -0.012$\pm$0.026 &  0.019$\pm$0.038 &       \\
$1313-333$ & 13 \ 16 \ 07.985935 & $-$33\ 38\ 59.17238 & 1.21 &  QSO & 0.10 & 0.13 &  0.056$\pm$0.028 & -0.033$\pm$0.034 & 2     \\
$1315+346$ & 13 \ 17 \ 36.494182 & $+$34\ 25\ 15.93251 & 1.05 &  QSO & 0.21 & 0.22 & -0.085$\pm$0.042 &  0.079$\pm$0.044 & 2     \\
$1324+224$ & 13 \ 27 \ 00.861312 & $+$22\ 10\ 50.16301 & 1.40 &  QSO & 0.10 & 0.13 & -0.038$\pm$0.020 &  0.021$\pm$0.025 &       \\
$1342+662$ & 13 \ 43 \ 45.959538 & $+$66\ 02\ 25.74508 & 0.77 &  QSO & 0.26 & 0.25 & -0.012$\pm$0.050 & -0.078$\pm$0.044 &       \\
$1347+539$ & 13 \ 49 \ 34.656626 & $+$53\ 41\ 17.04026 & 0.98 &  QSO & 0.16 & 0.11 & -0.033$\pm$0.031 &  0.010$\pm$0.020 &       \\
$1354+195$ & 13 \ 57 \ 04.436657 & $+$19\ 19\ 07.37233 & 0.72 &  QSO & 0.09 & 0.08 & -0.031$\pm$0.016 &  0.048$\pm$0.014 & 3     \\
$1354-152$ & 13 \ 57 \ 11.244969 & $-$15\ 27\ 28.78643 & 1.89 &  QSO & 0.09 & 0.11 &  0.048$\pm$0.018 & -0.021$\pm$0.018 & 2     \\
$1406-076$ & 14 \ 08 \ 56.481197 & $-$07\ 52\ 26.66628 & 1.49 &  QSO & 0.09 & 0.15 &  0.018$\pm$0.020 & -0.010$\pm$0.029 &       \\
$1448+762$ & 14 \ 48 \ 28.778906 & $+$76\ 01\ 11.59717 & 0.90 &  AGN & 0.13 & 0.11 & -0.053$\pm$0.032 &  0.006$\pm$0.028 &       \\
$1519-273$ & 15 \ 22 \ 37.675993 & $-$27\ 30\ 10.78539 & 1.30 &  BLL & 0.05 & 0.06 & -0.003$\pm$0.010 & -0.013$\pm$0.012 &       \\
$1606+106$ & 16 \ 08 \ 46.203179 & $+$10\ 29\ 07.77582 & 1.23 &  QSO & 0.01 & 0.01 &  0.007$\pm$0.002 &  0.006$\pm$0.003 & 3     \\
$1610-771$ & 16 \ 17 \ 49.276354 & $-$77\ 17\ 18.46743 & 1.71 &  QSO & 0.10 & 0.12 &  0.036$\pm$0.022 &  0.008$\pm$0.026 &       \\
$1622-297$ & 16 \ 26 \ 06.020829 & $-$29\ 51\ 26.97082 & 0.82 &  QSO & 0.16 & 0.16 &  0.014$\pm$0.042 &  0.023$\pm$0.042 &       \\
$1624+416$ & 16 \ 25 \ 57.669701 & $+$41\ 34\ 40.62921 & 2.55 &  QSO & 0.14 & 0.09 & -0.030$\pm$0.032 & -0.034$\pm$0.022 &       \\
$1633+382$ & 16 \ 35 \ 15.492974 & $+$38\ 08\ 04.50059 & 1.81 &  QSO & 0.08 & 0.09 &  0.000$\pm$0.008 &  0.008$\pm$0.010 &       \\
$1637+574$ & 16 \ 38 \ 13.456293 & $+$57\ 20\ 23.97916 & 0.75 &  QSO & 0.07 & 0.07 & -0.004$\pm$0.011 & -0.015$\pm$0.010 &       \\
$1638+398$ & 16 \ 40 \ 29.632774 & $+$39\ 46\ 46.02852 & 1.67 &  QSO & 0.01 & 0.02 & -0.008$\pm$0.003 & -0.003$\pm$0.004 & 2     \\
$1642+690$ & 16 \ 42 \ 07.848516 & $+$68\ 56\ 39.75643 & 0.75 &  AGN & 0.06 & 0.06 & -0.034$\pm$0.014 & -0.012$\pm$0.014 & 2     \\
$1656+053$ & 16 \ 58 \ 33.447346 & $+$05\ 15\ 16.44421 & 0.88 &  QSO & 0.14 & 0.18 & -0.046$\pm$0.026 & -0.057$\pm$0.032 &       \\
$1705+456$ & 17 \ 07 \ 17.753408 & $+$45\ 36\ 10.55269 & 0.65 &  QSO & 0.20 & 0.15 & -0.008$\pm$0.035 &  0.024$\pm$0.026 &       \\
$1726+455$ & 17 \ 27 \ 27.650808 & $+$45\ 30\ 39.73138 & 0.71 &  QSO & 0.01 & 0.01 & -0.011$\pm$0.003 &  0.006$\pm$0.004 & 3     \\
$1739+522$ & 17 \ 40 \ 36.977848 & $+$52\ 11\ 43.40750 & 1.38 &  QSO & 0.01 & 0.01 & -0.004$\pm$0.002 & -0.004$\pm$0.002 & 2     \\
$1741-038$ & 17 \ 43 \ 58.856140 & $-$03\ 50\ 04.61670 & 1.06 &  QSO & 0.01 & 0.01 & -0.010$\pm$0.002 &  0.000$\pm$0.003 & 3     \\
$1746+470$ & 17 \ 47 \ 26.647296 & $+$46\ 58\ 50.92627 & 1.48 &  BLL & 0.12 & 0.14 & -0.015$\pm$0.022 &  0.030$\pm$0.026 &       \\
$1749+701$ & 17 \ 48 \ 32.840244 & $+$70\ 05\ 50.76878 & 0.77 &  BLL & 0.13 & 0.08 & -0.016$\pm$0.026 & -0.004$\pm$0.016 &       \\
$1751+441$ & 17 \ 53 \ 22.647898 & $+$44\ 09\ 45.68608 & 0.87 &  QSO & 0.22 & 0.25 & -0.017$\pm$0.040 &  0.026$\pm$0.044 &       \\
$1758+388$ & 18 \ 00 \ 24.765361 & $+$38\ 48\ 30.69765 & 2.09 &  QSO & 0.07 & 0.10 &  0.002$\pm$0.015 & -0.029$\pm$0.022 &       \\
$1800+440$ & 18 \ 01 \ 32.314848 & $+$44\ 04\ 21.90032 & 0.66 &  QSO & 0.18 & 0.13 &  0.013$\pm$0.036 & -0.041$\pm$0.025 &       \\
$1803+784$ & 18 \ 00 \ 45.683913 & $+$78\ 28\ 04.01853 & 0.68 &  QSO & 0.01 & 0.01 & -0.001$\pm$0.002 &  0.005$\pm$0.002 & 3     \\
$1821+107$ & 18 \ 24 \ 02.855265 & $+$10\ 44\ 23.77392 & 1.36 &  QSO & 0.09 & 0.20 & -0.016$\pm$0.014 &  0.059$\pm$0.032 &       \\
$1823+568$ & 18 \ 24 \ 07.068374 & $+$56\ 51\ 01.49087 & 0.66 &  QSO & 0.03 & 0.04 & -0.006$\pm$0.011 & -0.017$\pm$0.013 &       \\
$1849+670$ & 18 \ 49 \ 16.072298 & $+$67\ 05\ 41.67999 & 0.66 &  QSO & 0.04 & 0.03 & -0.015$\pm$0.014 &  0.018$\pm$0.012 &       \\
$1901+319$ & 19 \ 02 \ 55.938889 & $+$31\ 59\ 41.70208 & 0.64 &  QSO & 0.09 & 0.09 &  0.038$\pm$0.020 & -0.042$\pm$0.018 & 2     \\
$1908-201$ & 19 \ 11 \ 09.652870 & $-$20\ 06\ 55.10864 & 1.12 &  QSO & 0.03 & 0.05 &  0.008$\pm$0.010 &  0.016$\pm$0.015 &       \\
$1936-155$ & 19 \ 39 \ 26.657728 & $-$15\ 25\ 43.05799 & 1.66 &  QSO & 0.08 & 0.11 &  0.022$\pm$0.026 & -0.049$\pm$0.034 &       \\
$1954-388$ & 19 \ 57 \ 59.819271 & $-$38\ 45\ 06.35625 & 0.63 &  QSO & 0.05 & 0.07 & -0.005$\pm$0.010 & -0.023$\pm$0.013 &       \\
$1958-179$ & 20 \ 00 \ 57.090449 & $-$17\ 48\ 57.67240 & 0.65 &  QSO & 0.03 & 0.04 & -0.004$\pm$0.005 & -0.029$\pm$0.007 & 3     \\
$2017+745$ & 20 \ 17 \ 13.079305 & $+$74\ 40\ 47.99994 & 2.19 &  QSO & 0.13 & 0.14 &  0.016$\pm$0.025 & -0.023$\pm$0.026 &       \\
$2029+121$ & 20 \ 31 \ 54.994277 & $+$12\ 19\ 41.34038 & 1.22 &  QSO & 0.09 & 0.15 &  0.002$\pm$0.017 & -0.009$\pm$0.026 &       \\
$2052-474$ & 20 \ 56 \ 16.359842 & $-$47\ 14\ 47.62771 & 1.49 &  QSO & 0.16 & 0.20 &  0.031$\pm$0.023 & -0.005$\pm$0.028 &       \\
$2121+053$ & 21 \ 23 \ 44.517384 & $+$05\ 35\ 22.09318 & 1.94 &  QSO & 0.02 & 0.03 &  0.019$\pm$0.004 & -0.007$\pm$0.005 & 3     \\
$2134+004$ & 21 \ 36 \ 38.586306 & $+$00\ 41\ 54.21344 & 1.93 &  QSO & 0.08 & 0.08 &  0.028$\pm$0.008 & -0.054$\pm$0.009 & 3     \\
$2136+141$ & 21 \ 39 \ 01.309267 & $+$14\ 23\ 35.99201 & 2.43 &  QSO & 0.02 & 0.03 & -0.005$\pm$0.005 & -0.020$\pm$0.007 & 2     \\
$2143-156$ & 21 \ 46 \ 22.979339 & $-$15\ 25\ 43.88518 & 0.70 &  QSO & 0.20 & 0.18 &  0.091$\pm$0.040 & -0.097$\pm$0.033 & 2     \\
$2145+067$ & 21 \ 48 \ 05.458677 & $+$06\ 57\ 38.60420 & 1.00 &  QSO & 0.01 & 0.02 & -0.031$\pm$0.003 &  0.014$\pm$0.004 & 3     \\
$2149+056$ & 21 \ 51 \ 37.875500 & $+$05\ 52\ 12.95458 & 0.74 &  QSO & 0.09 & 0.13 &  0.002$\pm$0.018 & -0.005$\pm$0.025 &       \\
$2209+236$ & 22 \ 12 \ 05.966316 & $+$23\ 55\ 40.54387 & 1.13 &  QSO & 0.05 & 0.09 &  0.004$\pm$0.010 & -0.017$\pm$0.017 &       \\
$2216-038$ & 22 \ 18 \ 52.037725 & $-$03\ 35\ 36.87944 & 0.90 &  QSO & 0.08 & 0.09 &  0.008$\pm$0.010 & -0.005$\pm$0.011 &       \\
$2223-052$ & 22 \ 25 \ 47.259293 & $-$04\ 57\ 01.39055 & 1.40 &  QSO & 0.02 & 0.03 &  0.006$\pm$0.006 & -0.026$\pm$0.008 & 3     \\
$2227-088$ & 22 \ 29 \ 40.084340 & $-$08\ 32\ 54.43534 & 1.56 &  QSO & 0.21 & 0.28 &  0.038$\pm$0.058 & -0.065$\pm$0.078 &       \\
$2234+282$ & 22 \ 36 \ 22.470865 & $+$28\ 28\ 57.41334 & 0.80 &  QSO & 0.01 & 0.02 & -0.025$\pm$0.002 & -0.001$\pm$0.003 & 3     \\
$2243-123$ & 22 \ 46 \ 18.231970 & $-$12\ 06\ 51.27693 & 0.63 &  QSO & 0.03 & 0.04 &  0.008$\pm$0.006 &  0.004$\pm$0.008 &       \\
\hline                       
\end{tabular} 
\end{minipage}
\end{table*}

\addtocounter{table}{-1}
\begin{table*}
\begin{minipage}{150mm}
\caption{continued}
\begin{tabular}{@{}lrrccrrrrl}
\hline 
ICRF     & \multicolumn{1}{c}{$\alpha_{2000}$} & \multicolumn{1}{c}{$\delta_{2000}$}&   Z&   src  & $\varepsilon\Delta\alpha$ & $\varepsilon\Delta\delta$ & \multicolumn{1}{c}{$\mu\alpha$}    & \multicolumn{1}{c}{$\mu\delta$}    & flag\\
\hline 
$2251+158$ & 22 \ 53 \ 57.747943 & $+$16\ 08\ 53.56094 & 0.86 &  QSO & 0.18 & 0.22 &  0.039$\pm$0.014 &  0.038$\pm$0.018 & 2     \\
$2252-090$ & 22 \ 55 \ 04.239779 & $-$08\ 44\ 04.02146 & 0.61 &  QSO & 0.24 & 0.35 &  0.069$\pm$0.069 & -0.115$\pm$0.085 &       \\
$2253+417$ & 22 \ 55 \ 36.707843 & $+$42\ 02\ 52.53261 & 1.48 &  QSO & 0.10 & 0.20 & -0.015$\pm$0.018 & -0.036$\pm$0.036 &       \\
$2255-282$ & 22 \ 58 \ 05.962888 & $-$27\ 58\ 21.25662 & 0.93 &  QSO & 0.04 & 0.05 &  0.010$\pm$0.007 & -0.019$\pm$0.010 &       \\
$2318+049$ & 23 \ 20 \ 44.856614 & $+$05\ 13\ 49.95246 & 0.62 &  QSO & 0.03 & 0.06 &  0.003$\pm$0.007 & -0.028$\pm$0.012 & 2     \\
$2320-035$ & 23 \ 23 \ 31.953751 & $-$03\ 17\ 05.02362 & 1.41 &  QSO & 0.21 & 0.34 & -0.014$\pm$0.095 &  0.140$\pm$0.145 &       \\
$2335-027$ & 23 \ 37 \ 57.339081 & $-$02\ 30\ 57.62928 & 1.07 &  QSO & 0.14 & 0.20 & -0.026$\pm$0.028 &  0.032$\pm$0.036 &       \\
$2351+456$ & 23 \ 54 \ 21.680267 & $+$45\ 53\ 04.23665 & 1.99 &  QSO & 0.13 & 0.16 & -0.020$\pm$0.022 & -0.017$\pm$0.028 &       \\
$2355-106$ & 23 \ 58 \ 10.882413 & $-$10\ 20\ 08.61133 & 1.62 &  QSO & 0.11 & 0.10 & -0.008$\pm$0.020 &  0.005$\pm$0.019 &       \\
$2356+385$ & 23 \ 59 \ 33.180781 & $+$38\ 50\ 42.31802 & 2.70 &  QSO & 0.03 & 0.04 &  0.000$\pm$0.007 & -0.015$\pm$0.007 & 2     \\
\hline                        
\end{tabular}
{\bf Table explanations}

1. Errors of coordinates and angular velocities are taken in milliarcseconds and in milliarcseconds per year respectively.

2. The last column contains flag which shows the confidence level (in rms) of source motion.

\end{minipage}
\end{table*}


\begin{thebibliography}{99}

\bibitem[\protect\citeauthoryear{Assafin et al.}{2003}]{ass03}
Assafin M., Zacharias M., Rafferty T.J. et al., 2003, AJ, {\bf 125}, p.2728--2739

\bibitem[\protect\citeauthoryear{Assafin et al.}{2007}]{ass07}
Assafin M., Nedelcu D.A., Badescu O.,  Popescu P., Andrei A.H., Camargo J.I.B., da Silva Neto D.N., Vieira Martins R., 2007, A\&A,  {\bf 476}, 989--993

\bibitem[\protect\citeauthoryear{Begelman, Blandford, Rees}{1984}]{beg84}
Begelman M.C., Blandford R.D., Rees M.J., 1984, Rev. Mod. Phys {\bf 56}, 255

\bibitem[\protect\citeauthoryear{Belokurov, Evans}{2002}]{evan01}
Belokurov V.A., Evans N.W., 2002, MNRAS, {\bf 331}, 649--665

\bibitem[\protect\citeauthoryear{Blandford, K\"onigl}{1979}]{bla79}
Blandford R.D., K\"onigl A., 1979, AJ, {\bf 232}, 34

\bibitem[\protect\citeauthoryear{Davydov et al.,}{2008}]{cher08}
Davydov V.V., Esipov V.F., Cherepashchuk A.M.,  2008, Astronomy Reports, {\bf 52}, 487

\bibitem[\protect\citeauthoryear{Doroshkevich}{2009}]{dor09}
Doroshkevich A.G., 2009, private communication

\bibitem[\protect\citeauthoryear{Feissel-Vernier}{2006}]{fei06}
Feissel-Vernier M., 2006, {\it Assessing astrometric quality: stability of the VLBI-derived extragalactic celestial frame}, in The International Celestial Reference System and Frame -- ICRS Center Report for 2001-2004, eds. Souchay J. and Feissel-Vernier M. (IERS Technical Note 34). Frankfurt am Main, Verlag des Bundesamts f\"ur Kartographie und Geod\"asie, 137 p., ISBN 3-89888-802-9, pp.49--71

\bibitem[\protect\citeauthoryear{Hinshaw et al.}{2009}]{hin09}
Hinshaw G. et al., 2009, ApJS, {\bf 180}, 225--245

\bibitem[\protect\citeauthoryear{Jackson, Jannetta}{2006}]{jack06}
Jackson J.C., Jannetta A.L., 2006, J. of Cosmology and Astroparticle Physics, Is.11, 2

\bibitem[\protect\citeauthoryear{Kovalevsky}{1995}]{Kov}
Kovalevsky J., 1995 , Modern Astrometry. Springer--Verlag, Berlin, Heidelberg

\bibitem[\protect\citeauthoryear{Ma}{1998}]{ma98}
Ma C., Arias E.F., Eubanks T.M. et al., 1998, AJ, {\bf 116}, 516

\bibitem[\protect\citeauthoryear{MacMillan}{2003}]{mcm03}
MacMillan D.C., 2003, in Romney J.D., Reid M.J., eds, Future Directions in High Resolution Astronomy, (The 10th Anniversary of the VLBI ASP Conference series). astro-ph/0309826

\bibitem[MacMillan \& Ma(2007)]{mcmma07} 
MacMillan D.S., \& Ma C.\, 2007, J. of Geodesy, {\bf 81}, 443 

\bibitem[\protect\citeauthoryear{Sazhin}{1996}]{sazh}
Sazhin M.V., 1996, Astronomy Letters, {\bf 22}, 573

\bibitem[\protect\citeauthoryear{Sazhin et al.}{1998}]{sazh98}
Sazhin M.V., Zharov V.E., Volynkin A.V., Kalinina T.A., 1998, MNRAS, {\bf 300}, 287

\bibitem[\protect\citeauthoryear{Sazhin et al.}{2001}]{sazh01}
Sazhin M.V., Zharov V.E., Kalinina T.A., 2001, MNRAS, {\bf 323}, 952

\bibitem[\protect\citeauthoryear{da Silva Neto et al.}{2002}]{net02}
da Silva Neto D.N., Andrei A.H., Vieira Martins R., and Assafin M., 2002, AJ, {\bf 124}, 612--618

\bibitem[Theuns et al.(2002)]{Theuns2002} 
Theuns T., Bernardi M., Frieman J., Hewett P., Schaye J., Sheth R.K., \& Subbarao M., 2002, ApJ Lett., {\bf 574}, L111

\bibitem[\protect\citeauthoryear{Titov}{2008a}]{tit08}
Titov O., 2008, Proper motion of reference radio sources, arXiv:0804.1403 

\bibitem[\protect\citeauthoryear{Titov}{2008b}]{tit09}
Titov O., 2008, Systematic effects in apparent proper motion of radio sources, arXiv:0805.1099 

\bibitem[\protect\citeauthoryear{Weinberg}{1972}]{wein72}
Weinberg S., 1972, Gravitation and Cosmology. John Wiley and Sons Inc., New York--London--Sydney--Toronto

\bibitem[\protect\citeauthoryear{WMAP recommendation}{2008}]{WMAPreco}
WMAP recommendation \ \ \underline{http://lambda.gsfc.nasa.gov/product/map/dr3/parameters\_summary.cfm} 


\bibitem[\protect\citeauthoryear{Zacharias et al.}{1999}]{zach99}
Zacharias N., Zacharias M.I., Hall D.M. et al., 1999, AJ, {\bf 118}, 2511--2525


\bibitem[\protect\citeauthoryear{Zharov}{2006}]{zhar06}
Zharov V.E., 2006, Spherical Astronomy. Fryasino, Vek-2, 478 p.

\bibitem[\protect\citeauthoryear{Zharov}{2009}]{zhar09a}
Zharov V., 2009, in press

\bibitem[\protect\citeauthoryear{Zharov et al.}{2009}]{zhar09}
Zharov V.E., Sazhin M.V., Sementsov V.N., Kuimov K.V., Sazhina O.S., 2009, Astronomicheskii Zhurnal, {\bf 86}, N7, p. 1 (in russian)

\end{thebibliography}
\end{document}